\documentclass[12pt,titlepage]{article}

\usepackage{html,hyperref}
\usepackage[pdftex]{graphicx}

\begin{document}

\centerline{\Huge Imaging the cool gas, dust, star formation,}
\centerline{\Huge and AGN in the first galaxies}

\vskip 0.2in
\noindent C.L. Carilli\footnote{Contact author: 
ccarilli@nrao.edu; National Radio Astronomy Observatory, 
P.O.\ Box O, Socorro, NM, 87801}(NRAO), S. Myers (NRAO), 
P. Appleton (HSC), F. Bertoldi (Bonn),
A. Blain (Caltech), D. Dale (Wyoming),  X. Fan (Arizona), 
Y. Li (CfA), K. Menten(MPIfR), 
K. Nagamine (UNLV), D. Narayanan (CfA), A. Omont (IAP), 
M. Strauss (Princeton), Yoshi Taniguchi (Ehime), J. Wagg (NRAO),
F. Walter (MPIA), A. Wolfe (UCSD), A. Wootten (NRAO), M.S. Yun (UMass)

\date{Draft v2.1 2009-02-09\\[2ex]
%%%%%%%%%%%%%%%%%%%%%%%%%%%%%%%%%%%%%%%%%%%%
%% TITLE FIGURE
%%%%%%%%%%%%%%%%%%%%%%%%%%%%%%%%%%%%%%%%%%%%
}

\vskip 0.4in

\begin{figure}[h]
\begin{center}
\includegraphics[width=4in]{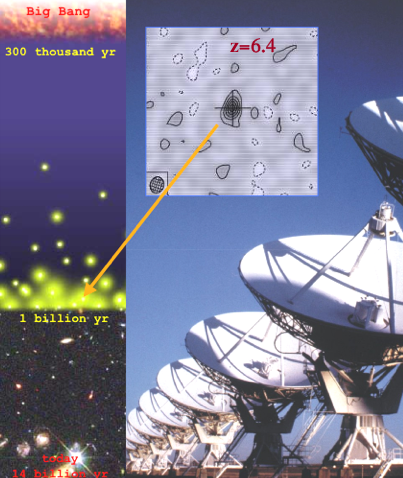}
\end{center}
\end{figure}

\clearpage\newpage

{\bf Abstract: When, and how, did the first galaxies and supermassive
black holes (SMBH) form, and how did they reionization the Universe?
First galaxy formation and cosmic reionization are among the
last frontiers in studies of cosmic structure formation. We delineate
the detailed astrophysical probes of early galaxy and SMBH formation
afforded by observations at centimeter through submillimeter
wavelengths.  These observations include studies of the molecular gas
(= the fuel for star formation in galaxies),  atomic fine structure
lines (= the dominant ISM gas coolant),  thermal dust continuum
emission (= an ideal star formation rate estimator), and radio
continuum emission from star formation and relativistic jets. High
resolution spectroscopic imaging can be used to study galaxy dynamics
and star formation on sub-kpc scales. These cm and mm observations are
the necessary compliment to near-IR observations, which probe the
stars and ionized gas, and X-ray observations, which reveal the
AGN. Together, a suite of revolutionary observatories planned for the
next decade from centimeter to X-ray wavelengths will provide the
requisite panchromatic view of the complex processes involved in the
formation of the first generation of galaxies and SMBHs, and cosmic
reionization.}

\vskip 0.2in

\bigskip
\leftline{\large{\bf 1 The formation of the first galaxies and SMBH}}
\medskip

Study of the formation of the first galaxies and super-massive black
holes is a key science driver for essentially all large area
telescopes, at all wavelengths, under construction or design.  Deep
near-IR surveys have revolutionized our understanding of early galaxy
formation by revealing star forming galaxies and SMBH back to the
near-edge of cosmic reionization, $z \sim 6$ to 7 (Ellis 2007, Saas Fe
Advanced Course 36 astroph/070124; Fan et al. 2006, ARAA, 44, 414).
Reionization sets a fundamental benchmark in cosmic structure
formation, corresponding to the epoch when the first luminous sources
(galaxies, quasars) reionize the neutral IGM.

Current cm and mm facilities are providing the first glimpse into the
cool gas, dust, and (obscuration-free) star formation in very early
(massive) galaxies.  In this white paper we describe the key
contributions that will be made in the next decade to the detailed
study of the first galaxies and SMBH with the powerful suite of large
centimeter (cm) and millimeter (mm) interferometers and single dish
telescopes.  We delineate a key science program on first galaxy
formation involving observations from cm through near-IR wavelengths,
along with the instrumentation development for the cm and mm
facilities that will leverage the major infrastructure investment for
dramatic science return.

\bigskip
\leftline{\large{\bf 2 Probing the era of first galaxies}}
\medskip

While progress in the field of very high $z$ galaxy formation has been
impressive, near-IR studies of the earliest galaxies are fundamentally
limited in two ways: {\it (i)} obscuration of rest-frame UV emission
by dust, and Ly$\alpha$ scattering in the neutral intergalactic
medium, may lead to a biased view of galaxy formation, and {\it (ii)}
near-IR studies reveal only the stars and ionized gas, thereby missing
the evolution of the cool gas in galaxies, the fuel for star
formation.  Line and continuum studies in the centimeter (cm) and
millimeter (mm) and submillimeter (submm) wavelength windows address both
these issues, by probing deep into the earliest, most active, and dust
obscured, phases of galaxy formation, and by revealing the molecular
and cool atomic gas.

\vskip 0.1in
\noindent\underline{\em Molecular gas:}
Molecular gas constitutes the fuel for star formation in galaxies, and
hence is a crucial probe of galaxy formation.  The key
physical diagnostics of such studies include:

\begin{list}{$\bullet$}{
\setlength{\topsep}{0ex}\setlength{\itemsep}{0ex plus0.2ex}
\setlength{\parsep}{0.5ex plus0.2ex minus0.1ex}}

\item CO rotational transitions: Emission from CO is the strongest of
the mm molecular lines from galaxies, and is the best tracer of the
total molecular gas mass, and of physical conditions via excitation
studies (Solomon \& vanden Bout 2007, ARAA, 43, 677).  Such studies
require observation of low to high order transitions, necessitating
observations at cm to mm wavelengths for high $z$ galaxies.

\item Dense gas tracers: Emission from high dipole moment molecules,
such as HCN and HCO+, trace the dense gas ($> 10^5$ cm$^{-3}$)
directly associated with star forming clouds (Gao et al. 2007, ApJ,
660, L93).  The higher order transitions may be sub-thermally excited
due to their very high critical densities ($> 10^7$ cm$^{-3}$),
accentuating the need for cm observations.

\item Gas dynamics: high resolution, spectroscopic imaging of the cool
gas is the most effective way to study the dynamics, and
dark matter content, of the first galaxies (Riechers et al. 2008, ApJ,
686, L9).  Such observations provide unique probes into key questions
on the evolution of the Tully-Fisher relation and  the black hole
mass to bulge mass relation.

\end{list}

\vskip 0.1in \noindent\underline{\em Fine structure lines (FSL):} Fine
structure line emission from galaxies provide critical diagnostics of
ISM physics and energetics (Stacey et al. 1991 ApJ, 373, 423),
including:

\begin{list}{$\bullet$}{
\setlength{\topsep}{0ex}\setlength{\itemsep}{0ex plus0.2ex}
\setlength{\parsep}{0.5ex plus0.2ex minus0.1ex}}

\item Emission lines from lower ionization species such as [CII]
158$\mu$m and [OI] 63, 145$\mu$m, are the dominant coolant in neutral
interstellar gas (Spitzer 1978, 'Physical Processes in the ISM'). The
[CI] 370 and 609 $\mu$m line ratio is a key ISM temperature probe
(Weiss et al. 2003, A\& A, 409, L41).

\item Higher ionization species such as [OIII] 52, 88 $\mu$m and
[NIII] 57 $\mu$m, trace the ionized ISM (Malhotra et al. 2001, ApJ
561, 766; Brauher et al. 2008 ApJS, 178, 280).

\item Studies of line ratios and their relationship to the FIR
luminosity and color can be used to derive gas temperature and
density, the interstellar radiation field (ISRF), and to constrain gas
heating and cooling (Brauher et al. 2008; Malhotra et al. 2001).

\end{list}

Extensive studies of FSL line emission in nearby galaxies, using
predominantly the [CII] 158 $\mu$m line, show a decrease of the
[CII]/FIR ratio with increasing FIR 'color temperature' (and FIR
luminosity), suggesting less efficient gas heating by photoelectron
ejection from dust grains due to positive grain charge induced by a
high UV ISRF (Malhotra et al. 2001).  However, these studies are
hampered by the low resolution available in the FIR, and hence
typically consider integrated properties of galaxies.  At high
redshift, these FSLs redshift into the 200--800~GHz submm bands and
can be imaged at sub-arcsecond resolution using ground-based interferometers.

A crucial research goal for understanding the high-redshift universe
is to determine how the observed neutral-gas in damped Lyman Alpha
absorption systems (DLAs) eventually fuels star formation.  Detection
of the [C II] 158$\mu$m in galaxies associated with DLAs is
crucial as it is an excellent tracer of neutral gas, is known to be
present as shown by the detection of CII* 1335.7 absorption in a
significant fraction of high-z DLAs, and, as the dominant gas coolant,
provides insight into FUV emission from young stars that heat the gas.

\vskip 0.1in

\noindent\underline{\em Submillimeter continuum:}
The well documented ``inverse K correction'' in the Rayleigh-Jeans 
submm spectrum of thermal emission from warm dust implies a
distance independent means of studying galaxies from $z \sim 0.5$ to
$z \sim 10$ (Blain et al. 2002, Phys. Rep. 369, 111). Imaging of this
thermal emission is vital to galaxy formation studies in
a number of ways:

\begin{list}{$\bullet$}{
\setlength{\topsep}{0ex}\setlength{\itemsep}{0ex plus0.2ex}
\setlength{\parsep}{0.5ex plus0.2ex minus0.1ex}}

\item Thermal emission from warm dust provides the one of cleanest method
for deriving total star formation rates in galaxies (Dale \& Helou 2002,
ApJ, 576, 159).

\item High resolution submm imaging reveals the distribution of
star formation, unhindered by obscuration.  For the first galaxies,
the peak in the FIR emission is shifted to the submm, allowing the use
of interferometers to perform sub-kpc resolution imaging. Combining
high resolution images of the rest-frame FIR emission with images of
the molecular gas presents a unique opportunity to study the evolution
of the Schmidt relation back to the epoch of reionization
(Krumholz \& Thompson 2007, ApJ, 669, 289).

\item Study of the dust SED yields a dust temperature and mass.

\end{list}

A particularly interesting problem is the formation of dust within 1
Gyr of the Big Bang, since the standard method of cool winds from AGB
stars takes too long. Numerous groups are considering alternate dust
formation mechanisms in the early Universe, likely associated with
massive star formation, eg. dust formation in primordial SNe (Maiolino
et al. 2004 Nature, 431, 533; Nozawa et al. 2009 arXiv:0812.1448).

\vskip 0.1in

\noindent\underline{\em Radio continuum:}
Radio continuum studies are routinely reaching detection limits of
$\sim 20$ to 30 $\mu$Jy at 1.4 GHz. Such studies probe
the first galaxies in a number of ways:

\begin{list}{$\bullet$}{
\setlength{\topsep}{0ex}\setlength{\itemsep}{0ex plus0.2ex}
\setlength{\parsep}{0.5ex plus0.2ex minus0.1ex}}

\item Radio emission from relativistic jets in AGN can be imaged to
milliarcsec resolution using VLBI (Momjian et al. 2008, AJ, 136, 344). Jets
likely play an important role in galaxy formation through hydrodynamic
feedback on the ISM (Zirm et al. 2005, ApJ, 630, 68).

\item The synchrotron continuum emission  provides an
obscuration-free measure of the massive star formation rate, via the
well-quantified radio-FIR correlation for star forming galaxies
(Condon 1992). Alternatively, deep cm and mm data can be used to test
the evolution of this relation back to the first galaxies.

\end{list}

%%%%%%%%%%%%%%%%%%%%%%%%%%%%%%%%%%%%%%%%%%%%
%% Figure:
\begin{figure}[t!]
\begin{center}
\includegraphics[width=6.5in]{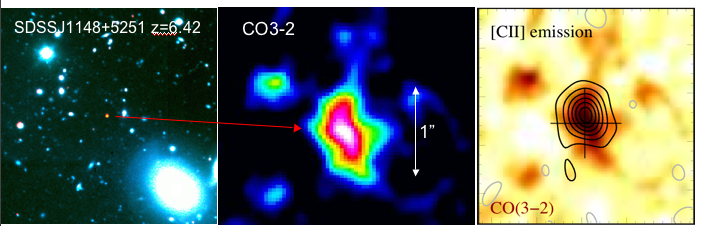}
%\vskip 2.5in
\caption{\label{fig:1}\small Images of SDSS J1148+5251 at $z = 6.42$. Left is 
a Keck true-color image (Djorgovski, Mahabal, and Bogosavljevic priv. comm.). 
Center is the VLA image of CO 3-2 emission (Walter et al. 2004).
Right is the PdBI [CII] 158 $\mu$m image (Walter et al. 2009). }
\end{center}
\end{figure}
%%
%%%%%%%%%%%%%%%%%%%%%%%%%%%%%%%%%%%%%%%%%%%%

\bigskip
\leftline{\large{\bf 3 The First Gyr --- coeval formation of galaxies and SMBH}}
\medskip

The power of submm, mm and cm observations to probe the most distant
galaxies is demonstrated by recent results on the host galaxy of the
most distant SDSS quasar, J1148+5251, at $z = 6.42$. The SDSS
observations, Keck spectroscopy, and HST imaging (White et al. 2005,
AJ, 129, 2102; White et al. 2003, AJ, 126, 1), reveal a SMBH of $\sim
2\times 10^9$ M$_\odot$. The host galaxy has been detected in thermal
dust, non-thermal radio continuum, CO line, and [CII] 158 $\mu$m
emission (Figure 1). High resolution imaging of the CO emission
reveals a massive reservoir of molecular gas, $2\times 10^{10}$
M$_\odot$, distributed on a scale of $\sim 6$ kpc in the host galaxy
(Walter et al. 2004).  The broad band SED of J1148+5251, shows a clear
FIR excess, consistent with 50K dust emission and with the radio-FIR
correlation for star forming galaxies (Wang et al. 2008, ApJ, 687,
848). The high CO excitation in J1148+5251 (Bertoldi et al. A\& A,
406, L55) is comparable to that seen in starburst nuclei implying a
predominantly dense ($\sim 10^5$ cm$^{-3}$), warm ISM.  Recent high
resolution imaging of the [CII] emission in the host galaxy of
J1148+5251 reveals an extreme starburst region with a diameter of
$\sim 1.5$kpc (Walter et al.  2009, Nature, 475, 699), forming stars
at the 'Eddington limited' rate of $\sim 1000$ M$_\odot$ year$^{-1}$
kpc$^{-2}$ (Thompson et al. ApJ, 630, 167). Such a high surface
density of star formation is seen in the starburst nuclei of low z
ULIRGs, such as Arp 220, although on a much smaller scale ($< 100$pc).

These results, and results on similar $z \sim 6$ quasar host galaxies,
are consistent with the co-eval formation of massive galaxies, and
SMBH, at the earliest epochs. Cosmological numerical simulations have
been used to elucidate such systems  (Li et al. 2007. ApJ, 665,
L187) who find that the SMBH forms through Eddington limited accretion and BH
mergers, while the host galaxy experiences vigorous star formation 
($\sim 10^3$ M$_\odot$ year$^{-1}$) during a series of major gas rich
mergers, starting at $z \sim 14$.  These systems are destined to
become giant elliptical galaxies at the centers of rich clusters.

%%%%%%%%%%%%%%%%%%%%%%%%%%%%%%%%%%%%%%%%%%%%
%% Figure:
\begin{figure}[t!]
\begin{center}
\includegraphics[width=6.5in]{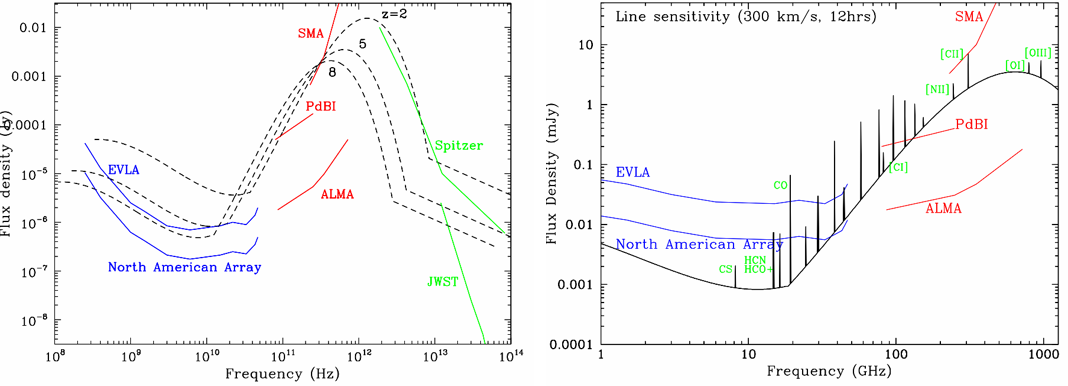}
%\vskip 2.5in
\caption{\label{fig:3}\small {\bf Left:} The solid curves show the 
(1$\sigma$) sensitivity to continuum emission for select current and 
future cm and mm interferometers, as well as for space IR telescopes,  
in 12 hours. The dashed
curves are the continuum spectra for Arp 220, at $z = 2$, 5,
and 8.  {\bf Right:} The colored curves show the (1$\sigma$)
sensitivity to spectral line emission in 12 hours for a 300 km
s$^{-1}$ line width.  The solid curve is the spectrum,
including molecular and atomic fine structure lines, for Arp 220 at $z
= 5$.  }
\end{center}
\end{figure}
%%
%%%%%%%%%%%%%%%%%%%%%%%%%%%%%%%%%%%%%%%%%%%%

\bigskip
\leftline{\large{\bf 4 Key Science Program: First Galaxies 2010--2020}}
\medskip

In this section, we outline a science program to find and exploit the
first galaxies.

\vskip 0.1in

\noindent\underline{\em First galaxy candidates:} Candidates $z > 6$
can be identified in deep, wide area surveys in the near-IR, radio,
and submm.  For example, near-IR deep survey instruments such as
Hyper-Suprime-Cam at Subaru and Vista at ESO will reveal thousands of
Lyman break (LBG) and Lyman alpha (LAE) galaxy candidates at $z>6$,
while wide field surveys such as Pan-STARRS and LSST will discover $z
\sim 7$ quasar candidates.  Multi-wavelength, large format bolometer
cameras on single dish mm telescopes such as the GBT, LMT, and CCAT,
along with deep EVLA radio surveys at 1.4 GHz, will identify similar
numbers of distant, dusty star forming galaxies, with redshift
estimates based on the broad-band SEDs.

\vskip 0.1in
\noindent\underline{\em Spectroscopic confirmation:}
Follow-up wide-band spectroscopy can then be done in the
mm and near-IR to determine spectroscopic redshifts and the global
properties of the galaxies.  As there will be large numbers of 
candidates fed in by the surveys, a key role will be played in these studies
by very wide-band spectrometers on large mm telescopes. 

\vskip 0.1in \noindent\underline{\em High Resolution Imaging and
spectroscopy:} Once identified, high resolution spectroscopic imaging
with ALMA and the EVLA will delineate the gas, dust, star formation,
and AGN at sub-kpc spatial resolution ($< 0.2"$), as described in the
sections above. In parallel, the JWST and TMT will provide
ultra-sensitive spectroscopy and imaging of the stars, ionized gas,
and AGN.

\vskip 0.1in

Figure 2a shows the continuum sensitivities of current and
future telescopes, along with the radio through near-IR SED of an
active star forming galaxy, Arp 220 (star formation rate $\sim 100$
M$_\odot$ year$^{-1}$) at $z = 2, 5$, and 8. Current telescopes, such
as Spitzer, the Plateau de Bure, and, soon, the EVLA, are able to
detect such active star forming galaxies into cosmic reionization. The
increased sensitivity of JWST and ALMA will push down to ``normal''
star forming galaxies, e.g. LAE and LBGs, with star formation rates
$\le 10$ M$_\odot$ year$^{-1}$.

Figure 2b shows the line intensity of Arp 220 at $z=5$, along with
the rms line sensitivity of current and future radio interferometers.
Centimeter telescopes study the low order molecular line transitions,
while in the mm, higher order molecular line transitions, as well as
the atomic FSL, are observed. 

\bigskip
\leftline{\large{\bf 5 Instrumenting for First Galaxy Science}}
\medskip

Galaxy formation is a complex process, and it is clear that a
panchromatic approach is required to understand the myriad processes
involved in early galaxy formation.  In this white paper, we call
particular attention to large radio, millimeter, and submillimeter
facilities that will play a crucial and complimentary role in the
field over the coming decade.

\vskip 0.1in \noindent\underline{\em Survey Telescopes:}
Large-aperture millimeter and submillimeter telescopes such as CCAT,
GBT, and LMT equipped with large wide-field bolometer cameras and
wide-band spectroscopic receivers will be able to carry out the key
surveys that form the basis of the follow-up detailed
studies. Ultra-sensitive, multiwavelength continuum surveys can be
used to identify candidate $z > 6$ galaxies, while very wide band
spectroscopy ($\ge 32$GHz) can be used to determine redshifts from
molecular and fine structure lines.

\vskip 0.1in 
\noindent\underline{\em ALMA:} The Atacama Large Millimeter Array
(ALMA)\footnote{http://www.almaobservatory.org/} will do  the
heavy-lifting for detailed studies of the cool gas, dust, and star
formation in early galaxies. ALMA will have close to two orders of
magnitude improved sensitivity, spatial resolution, and spectral
capabilities over existing (sub)mm interferometers.  As a concrete
example, a $z \sim 7$ galaxy with a $\sim 10$ M$_\odot$ year$^{-1}$
star formation rate has an expected peak [CII] 158 $\mu$m line
(redshifted to 240 GHz) flux density $\sim 0.5$mJy and width $\sim
200$ km s$^{-1}$, which could be detected by ALMA at $5\sigma$ in 1
hour.  Longer exposures can be used for detailed imaging at sub-kpc
resolution, as well as to look for other key diagnostic FSL, such as
redshifted [OI] 63$\mu$m and [OIII] 52 $\mu$m lines.
Figure 3 shows the predicted 93GHz ALMA spectrum of
J1148+5251 at $z = 6.42$. Future large area mm telescopes allow for
true multi-line spectroscopy, comparable to optical spectra.

%%%%%%%%%%%%%%%%%%%%%%%%%%%%%%%%%%%%%%%%%%%%
%% Figure:
\begin{figure}[t!]  
\begin{center}
\includegraphics[width=5in]{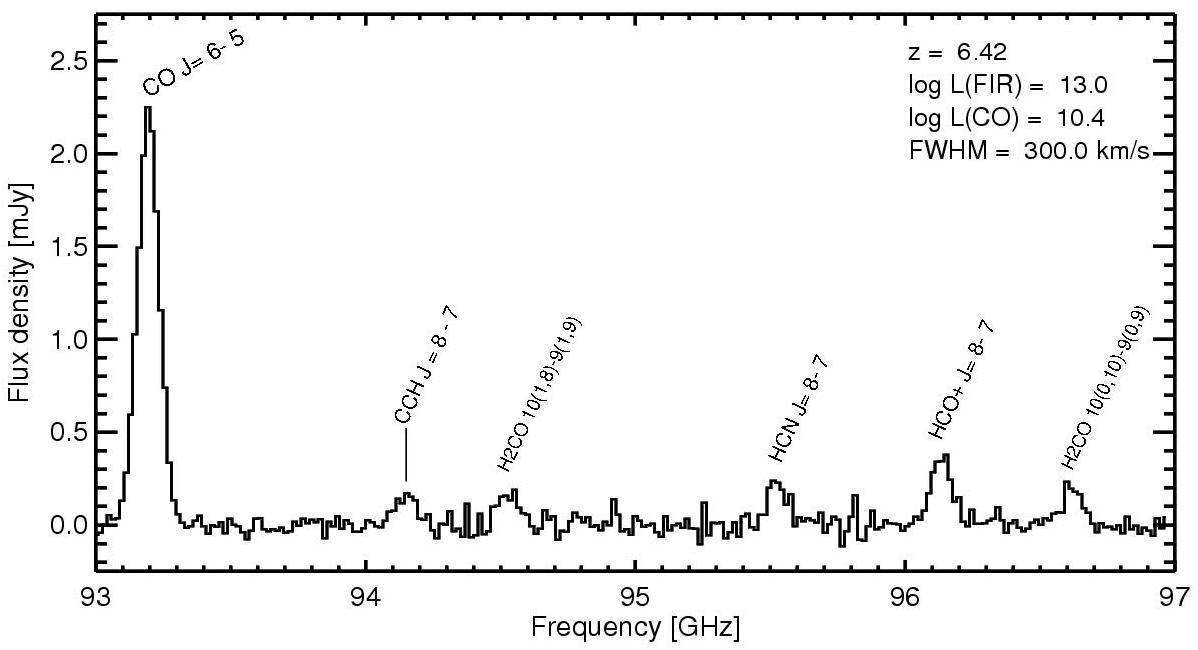}
%\vskip 2.5in
\caption{\label{fig:2}\small A simulated spectrum at 93GHz of
J1148+5251 ($z = 6.42$) for a 24hr exposure on ALMA. This is only one
of the two 4 GHz sidebands.}  
\end{center}
\end{figure}
%%
%%%%%%%%%%%%%%%%%%%%%%%%%%%%%%%%%%%%%%%%%%%%

There are a number of development projects for ALMA that are being
considered that will provide dramatic science return for modest
investments, leveraged on the existing infrastructure.  These include:
(1) completion of the ALMA receiver bands beyond those funded in the
construction project, such as 165--210~GHz (ALMA band 5) covering the
key ISM cooling [CII] 158$\mu$m line for $8 < z < 10.6$, and the
Terahertz bands (Bands 10 and 11: 750 to 950GHz, and 1240 to 1520 GHz,
respectively) for the lowest order rotational transitions from H$_2$
at $z > 7$ (rest wavelength 28$\mu$m).  (2) A second generation
correlator for wide-band spectroscopy with 32~GHz bandwidth, improving
capabilities for redshift determinations, astrochemistry, and
continuum sensitivity.  (3) Wide-field upgrades using focal plane
arrays to increase survey speed dramatically.

\vskip 0.1in

\noindent\underline{\em EVLA:} The Expanded Very Large Array
(EVLA)\footnote{http://www.aoc.nrao.edu/evla/} is a cornerstone
of the program for the detailed study of first galaxies.  The EVLA
upgrade provides complete frequency coverage from 1--50~GHz with 8~GHz
of instantaneous bandwidth and $10^4$ spectral channels, giving an
order of magnitude improvement in continuum sensitivity at the upper
end.  The EVLA will image the low order molecular lines in the
20--50~GHz bands at sub-arcsecond resolution, and provide
unprecedented sensitivity to the radio continuum emission at 1--4~GHz,
at arcsecond resolution.  Deep and wide field observations will image
to $\sim 1$ $\mu$Jy levels , adequate to detect active star formation
($\ge 50$ M$_\odot$ year$^{-1}$) at $z \sim 6$, with stacking analyses
potentially pushing an order of magnitude further for statistical
studies.

We note a few of the key EVLA development projects proposed for
the coming decade that will greatly benefit studies of first light: 
(1) A state-of-the-art low frequency system (75 to 350 MHz) would
enhance the ability of the EVLA to image steep spectrum (AGN
and starburst) radio emission with the resolution (a few
arcseconds) needed to avoiding confusion.   
(2) Real-time interferometric phase correction via water vapor
radiometry is being planned to increase the efficiency of observing at
the highest frequencies at the EVLA. 
(3) The 'North American Array' is a technology development program with
the long-term goal of increasing the EVLA collecting area and extending 
baselines to a few hundred kilometers. Such improvements are required to 
image thermal emission and study the cool molecular gas reservoirs in normal
galaxies at the highest redshifts.  This is the first step towards a
high-frequency component of the Square Kilometer
Array\footnote{http://www.skatelescope.org} program, targeted at
exploring the sub-$\mu$Jy sky in the decade beyond 2020.

\bigskip
\leftline{\large{\bf 6. Conclusions}}
\medskip

Cosmic reionization and the formation of the first galaxies and SMBH
is among the last frontiers in the study of cosmic structure
formation.  We envision a clear path for the discovery, and detailed
study, of the first galaxies afforded by a powerful suite of
telescopes in the next decade.  Near-IR telescopes (eg. JWST, TMT)
will reveal the stars and ionized gas, while X-ray telescopes (eg. the
IXO) will study the AGN. Far-IR telescopes (eg. SPICA, CALISTO)
provide views of the rest-frame mid-IR emission, including the
molecular hydrogen rotational transitions.  The submm, mm, and cm-wave
observations discussed in this white paper probe the dust and cool
gas, the fuel for star formation, provide an unobscured view of star
formation and AGN, and constrain galaxy dynamics on kpc-scales.  In
parallel, low frequency 'redshifted HI 21cm cosmology' telescopes will
image the evolution of the neutral IGM on large scales.  Together,
this suite of revolutionary observatories will provide the requisite
panchromatic view of the complex processes involved in cosmic
reionization and the formation of the first generation of galaxies and
SMBHs.

\end{document}